\documentclass{aa}

\usepackage{natbib}

\usepackage{txfonts}
\usepackage{hyperref}

\usepackage{graphicx}   
\usepackage{multirow} 
\usepackage{makecell} 

\newcommand{\be}{\begin{equation}}
\newcommand{\ee}{\end{equation}}
\newcommand{\bea}{\begin{eqnarray}}
\newcommand{\eea}{\end{eqnarray}}

\newcommand{\unit}{\,\mathrm}

\DeclareMathAlphabet\mathbfcal{OMS}{cmsy}{b}{n}

\hypersetup{colorlinks=true, citecolor=blue, linkcolor=blue}

\usepackage{color}

\begin{document}

\title{Circumbinary planets: migration, trapping in mean-motion resonances, and ejection}

   \author{Emmanuel Gianuzzi
          \inst{1,2,3}\fnmsep\thanks{E-mail: \href{mailto:egianuzzi@mi.unc.edu.ar}{egianuzzi@mi.unc.edu.ar}},
          Cristian Giuppone\inst{2,3},
          Nicolás Cuello\inst{4}
          }

   \institute{
   Facultad de Matem\'atica, Astronom\'ia, F\'isica y Computaci\'on (FAMAF), Universidad Nacional de C\'ordoba (UNC), C\'ordoba, Argentina,
   \and
   Instituto de Astronom\'ia Te\'orica y Experimental (IATE), CONICET-UNC, C\'ordoba, Argentina,
   \and
   Observatorio Astron\'omico de C\'ordoba (OAC), UNC, C\'ordoba, Argentina,
   \and
   Univ. Grenoble Alpes, CNRS, IPAG / UMR 5274, F-38000 Grenoble, France.
   }

\date{}

\abstract
{
Most of the planetary systems discovered around binary stars are located at approximately three semi-major axes from the barycentre of their system, curiously close to low-order mean-motion resonances (MMRs). The formation mechanism of these circumbinary planets is not yet fully understood. In situ formation is extremely challenging because of the strong interaction with the binary.
One possible explanation is that, after their formation, the interactions between these planets and the surrounding protoplanetary disc cause them to migrate at velocities dependent on the nature of the disc and the mass of the exoplanet. Although extensive data can be obtained with direct hydrodynamical simulations, their computational cost remains too high. On the other hand, the direct n-body simulations approach allows us to model a large variety of parameters at much lower cost.
}
{
We analyse the planetary migration around a wide variety of binary stars using Stokes-like forces that mimic planetary migration at a constant rate. Our goal is to identify the main parameters responsible for the ejection of planets at different resonances with the inner binary.
}
{
  We performed 4200 n-body simulations with Stokes-like forces and analysed their evolution and outcome as a function of the properties of each system. For each simulated exoplanet, we applied an ensemble learning method for classification in order to clarify the relationship between the inspected parameters and the process of MMR capture.
} 
{
We identify the capture probability for different N/1 MMRs, 4/1 being the most prone to capture exoplanets, with $37\%$ probability, followed by MMR 5/1 with $\sim23\%$ of probability. The eccentricity of the binary is found to be the most important parameter in determining the MMR capture of each circumbinary exoplanet, followed by the mass ratio of the binary and the initial eccentricity of the planet.}
{}


\titlerunning{Circumbinary planets: migration, trapping in mean-motion resonances, and ejection}

\authorrunning{Gianuzzi et al.}
\maketitle

\section{Introduction}\label{sec:intro}
Circumbinary planets (CBPs) discovered around transiting binaries challenge our knowledge of planet formation and migration in the disc around binaries. There are 13 transiting CBPs orbiting 11 Kepler-eclipsing binaries. Recently, the first circumbinary planet was reported using data from the TESS mission \citep[TOI~1338,][]{Kostov2020}. Most of the planets lie near their instability boundary at about three to five binary separations in regions surrounded by strong resonances. Recent studies argue that the occurrence rate of giant Kepler-like CBPs is comparable to that of giant planets in single-star systems \citep[see][and references therein]{Martin2019}.

\renewcommand{\arraystretch}{0.9}
\begin{table}
\begin{tiny}
\begin{center}
\caption{Parameters of the 14 known transiting CBPs.}
\begin{tabular}{c|cccccccccccc}
\hline
Name     & $q$  & $e_{\rm bin}$  & $R_{\rm p}$ & $M_{\rm p}$    & Metallicity\\
         &      &                &($R_\oplus$) & ($M_{\oplus}$) & (dex)\\
\hline
K-16     & 0.29 & 0.16  & 8.27  &104.84     & -0.3 [m/H]\\
K-34     & 0.97 & 0.52  & 8.38  & 69.89     & -0.07 [m/H]\\
K-35     & 0.91 & 0.14  & 7.99  & 44.78     & -0.34 [m/H]\\
K-38     & 0.26 & 0.10  & 4.20  & $<122$    & -0.11 [m/H]\\
K-47b    & 0.35 & 0.023 & 3.05  & $<25.77$  & -0.25 [m/H]\\
K-47d    & 0.35 & 0.023 & 7.04  & 19.02     & -0.04 [M/H]\\
K-47c    & 0.35 & 0.023 & 4.65  & 3.17      & -0.25 [m/H]\\
K-64     & 0.27 & 0.21  & 6.10  & 168.70    & 0.21 [Fe/H] \\
K-413    & 0.66 & 0.037 & 4.35  & $<168.7$  & -0.2 [Fe/H]\\
K-453    & 0.20 & 0.051 & 6.30  & $< 15.88$ & 0.09 [m/H]\\
K-1647   & 0.80 & 0.16  & 11.64 & 483.00    & -0.14 [Fe/H]\\
K-1661   & 0.31 & 0.112 & 3.87  & 17.00     & -0.12 [M/H]\\
TOI-1338 & 0.29 & 0.156 & 6.90  & 30.20     & 0.01 [Fe/H]\\
TIC 1729 & 0.97 & 0.448 & 11.25 & 942       & 0.34 [Fe/H]\\
\hline
\end{tabular}
\tablefoot{\emph{K} stands for \emph{Kepler}, and TIC 17290098 is identified as TIC-1729. For TIC-1729, we only show the solution named \textit{Family 5} by the authors \citep{Kostov2021}. Discovery papers: 16 \citep{Doyle2011}; 34 \& 35 \citep{Welsh2012}; 38 \citep{Orosz2012b}; 47 \citep{Orosz2012a,Orosz2019}; 64 \citep{Schwamb2013,Kostov2013}; 413 \citep{Kostov2014}; 453 \citep{Welsh2015}; 1647 \citep{Kostov2016}; 1661; \citep{Socia2020}; 1338 \citep{Kostov2020}; 1729 \citep{Kostov2021}.}
\label{tab:known_planets}
\end{center}
\end{tiny}
\end{table}

Transiting binaries that host planets have short periods ($<30$ days) \citep{Schwarz+2016} with eccentricities ranging from quasicircular (such as \textit{Kepler}-47, \citealt{Orosz2012a}) to highly eccentric  orbits (such as \textit{Kepler}-34, \citealt{Welsh2012}). Typical planets detected around binary stellar systems have a radius of about $10$ Earth radii and orbital periods of about $160$ days on almost circular coplanar orbits (i.e. $a_{\rm p} \sim 0.35\unit{AU}$). Radius, period, and coplanarity values are affected by observational biases \citep{Martin2018}. However, long-period CBPs could very well exist, with various masses and sizes.

There is some evidence that the $N/1$ mean motion resonances (MMRs) are related to the planetary parking location and their evolution in the disc, as they lead to strong regions of chaos around the binaries \citep{Gallardo2021}. For example, \citet{Zoppetti2018} applied a simplified model to Kepler-38 and suggested a capture in 5/1 MMR with subsequent tidal evolution of the planet outside MMRs. Other authors have performed fine-tuning hydrodynamical simulations, placing some constraints on the protoplanetary discs that allow migration of CBPs, but these simulations do not show capture at MMRs \citep[see e.g.][and references therein]{Thun2018, Penzlin2019, Penzlin2020, Penzlin2021}. In all these simulations, the authors find that the final parking locations of CBPs depend on the binary orbital parameters, which affect the CBD morphology. Essentially, the planets migrate to the inner edge of the disc cavity and remain there. But this is a problem because the parking locations seem to be too large when compared to recent observations. In their pioneering work, \citet{Nelson2003} showed a temporary capture in MMR 4/1 around a binary. However, the planet was placed inside the disc gap and the system was integrated for a few hundred binary orbital periods only.

In situ planet formation at these close orbits around binaries is extremely unlikely. 
Strong perturbation forces close to the binary inhibit planetesimal and dust accretion \citep{Moriwaki2004, Paardekooper2012,Silsbee2015,Meschiari2012}, and turbulence induced by hydrodynamical parametric instabilities can dramatically reduce pebble accretion efficiencies \citep{Pierens2020}.
A more likely scenario is planet formation in the outer disc and subsequent disc-driven migration to the observed close orbits \citep{Pierens2008,Bromley2015}. However, this assumption does not explain the particular stopping positions of most CBPs. 

\begin{figure*}
\centering
\includegraphics[width=0.99\columnwidth, clip]{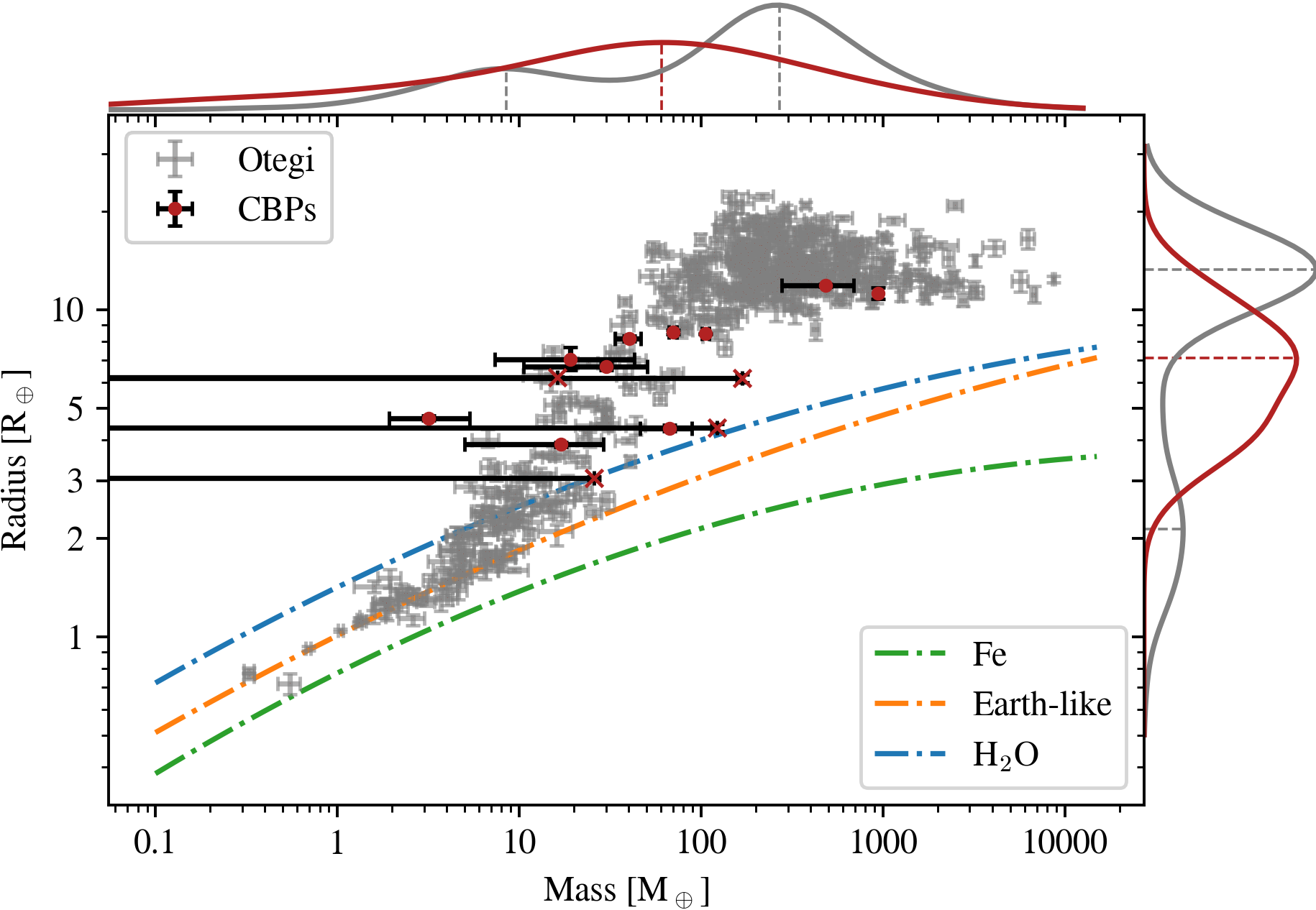}
\includegraphics[width=0.95\columnwidth, clip]{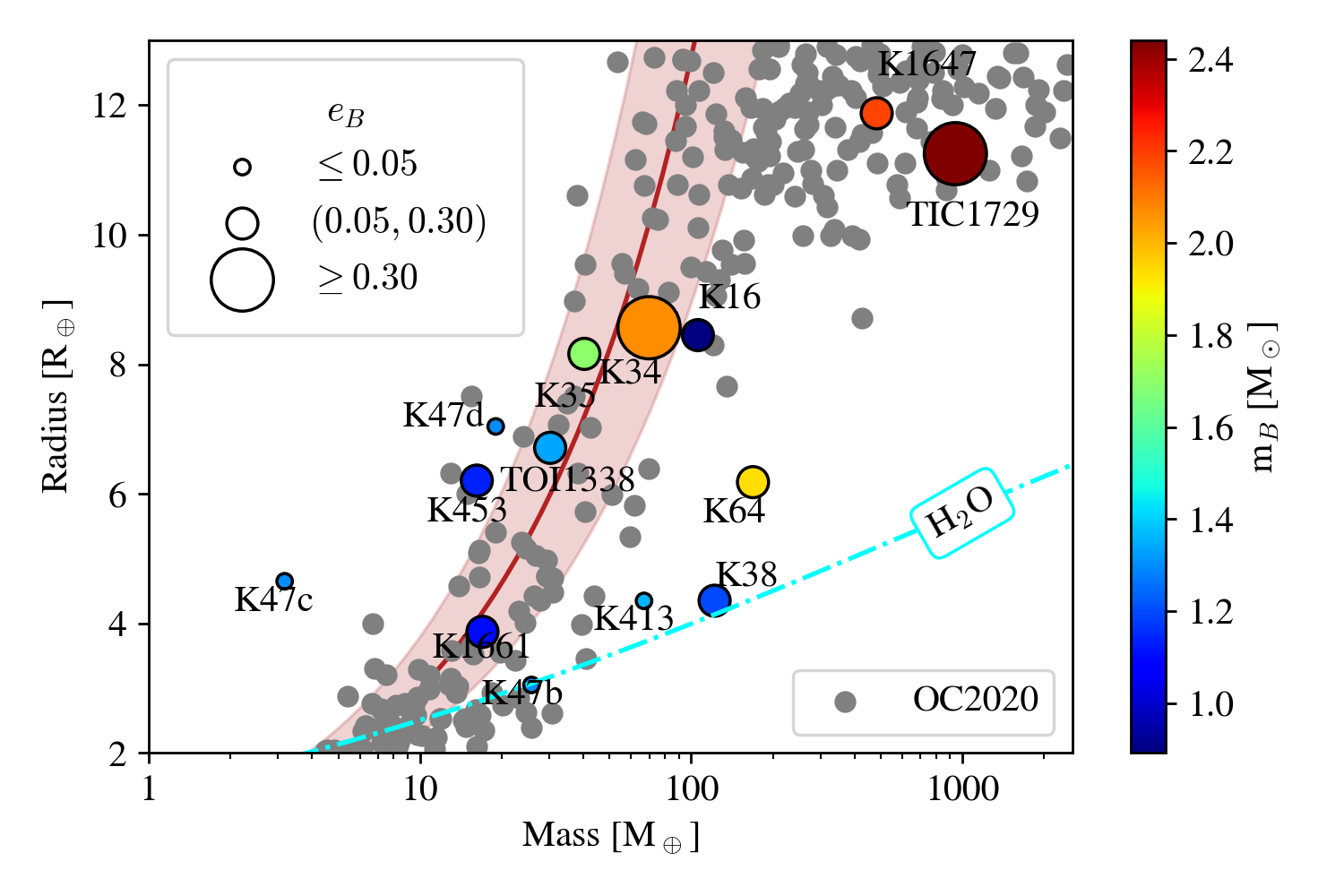}
\caption{Mass--radius relation of transiting circumbinary exoplanets and circumstellar exoplanets. Left panel: Radii of exoplanets presented in OC2020 (grey) and CBPs (red/black) as a function of their mass. Their respective error estimations are presented with error bars. In the case of $K38$, $K47b$, $K413$, and $K453,$ we mark their maximum mass reference value with cross symbols ($\times$).
The light-blue, orange, and green lines denote the theoretical $H_2O$, $Fe,$ and $Earth$-$like$ composition lines.
The curves outside the box represent the kernel density estimation of the mass (top) and radius (right) of the scattered data points. The data bandwidth for density estimation is calculated with Scott’s Rule \citep{Scott2015} using the {\tt scipy} Python package. The vertical (top) and horizontal (right) dashed lines denote the locations of the maximum values in each distribution. All distributions are normalised to their own total counts. Right panel: Zoom onto the CBP region shown in the left panel.
The exoplanets presented in OC2020 are shown as grey circles, while the CBPs are coloured according to their binary mass. The size of each CBP circle is proportional to the eccentricity of the binary. The red shaded area denotes the mass--radius relation (the red solid line being its reference values) presented in OC2020}\label{fig:m_r_Otegi}
\end{figure*}

In planet-forming discs, gas is partially supported against gravity by its pressure gradient. Therefore, the surrounding planetesimals experience aerodynamic drag, which translates into eccentricity damping. This mechanism is thought to increase their stability around single stars \citep{Kley2012, Sutherland2019}. However, the disc surrounding a binary may have a non-axisymmetrical distribution and become eccentric \citep{Kley2008, Ragusa+2017, Thun+2017, Poblete+2019, Hirsh+2020, Paardekooper2022arXiv} because of the effects of the secondary star. 

Several studies have already investigated planet migration in circumbinary discs \citep{Nelson2003, Pierens2008, Pierens2013, Kley2014, Kley2015, Kley2019}. Because of gravitational torques, the binary clears out the inner cavity and planet migration is stopped at the inner edge of the disc at a location close to the binary but not yet matching the observations in most cases studied.
Numerical simulations show that the inner disc shape and therefore the final planet orbital parameters depend on disc parameters such as viscosity and scale height, and the mass ratio and eccentricity of the binary \citep{Mutter2017,Thun2018, Ragusa2020}

Our aim is to study planetary migration around different kinds of binaries and to elucidate the possible physical parameters of discs that can explain the final parking position of the planets.
Taking into account the two exoplanets recently presented by \citet{Esmer2022}, 42 circumbinary exoplanets have so far been discovered\footnote{Obtained from \href{https://exoplanetarchive.ipac.caltech.edu/}{Nasa Exoplanet Archive} database}. We mainly rely on the findings of the Kepler transit survey when examining trends of circumbinary planets. This is because, in contrast to the many limitations of the suggested eclipse time variation planets, it is the only sample that is sufficiently large for preliminary population studies and also contains reliable discoveries. Furthermore, by restricting ourselves to a single observation method, only a single observation bias needs to be taken into consideration \citep{Martin2018}. In the remainder of this paper, we restrict our study to CBPs detected by transit methods only.

Table \ref{tab:known_planets} shows the planetary and star parameters from known transiting CBPs. We observe that typical planets around binaries are quite circular ($e_\text{mean}\simeq0.06$), which could be a consequence of the interaction with their natal protoplanetary disc. All the planets have radii of greater than $3R_{\oplus}$. Planet masses are typically poorly constrained or compatible with zero.
We take metallicities from discovery papers, because we find that the Exoplanet.eu database\footnote{\href{http://exoplanet.eu/catalog/}{http://exoplanet.eu/catalog/}} has missing values. For instance, the discovery paper of TOI-1338 \citep{Kostov2020} reports [Fe/H]$=-0.01\pm0.05\unit{dex}$ (measured with \emph{HARPS}) while its metallicity in the Exoplanet.eu database is [Fe/H]$=-0.4\pm0.1\unit{dex}$.

\begin{figure}
\centering
\includegraphics[width=1.\columnwidth, clip]{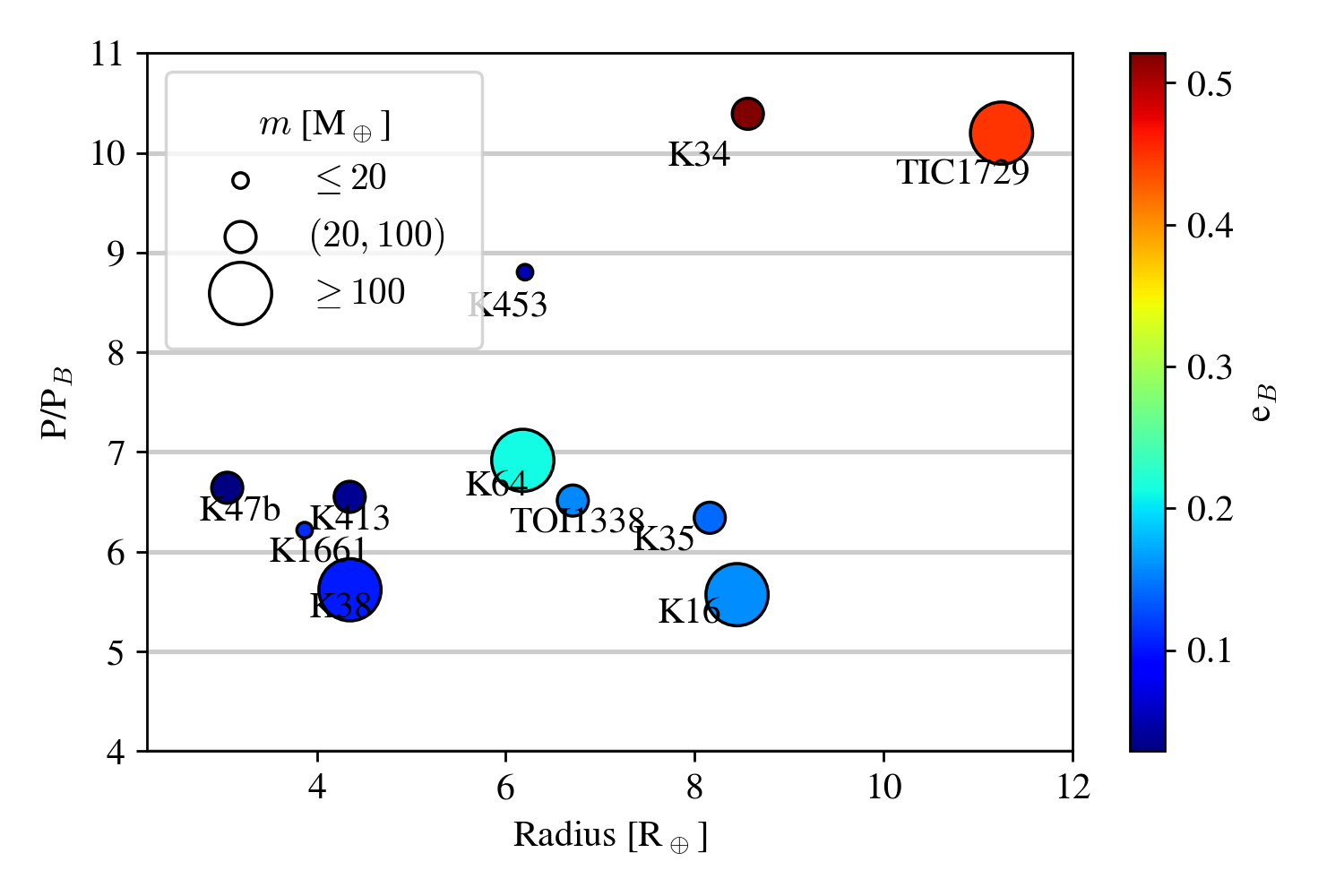}
\caption{Period ratio of known circumbinary planets as a function of their radii. The size of each point represents the planet mass, and the colour scale indicates its binary eccentricity. Horizontal grey lines denote the location of the resonances $N/1$. The plot is truncated to $P/P_B|_{\max}=11$.}
\label{fig:R.P_Pb}
\end{figure}

Figure \ref{fig:m_r_Otegi} shows the masses of exoplanets presented in the catalogue of \citet[][hereafter OC2020]{Otegi2020} as a function of their radius (grey markers). This catalogue contains data for planets with masses lower than $120\unit{M}_\oplus$, with reliable and robust mass and radius measurements. Known CBPs are also included (red markers) for comparison. The distribution of radii of exoplanets is bimodal \citep{David2021}, with peaks at $\sim2\unit{R}_\oplus$ and $\sim13\unit{R}_\oplus$, while the CBP distribution has a peak close to the valley of this distribution, at $\sim7\unit{R}_\oplus$. On the other hand, although the mass distribution of exoplanets can also be categorised as bimodal with peaks at $\sim9\unit{M}_\oplus$ and $\sim260\unit{M}_\oplus$, the mass distribution of CBPs is harder to categorise because of the large errors in their estimations. However, this distribution exhibits a peak close to the valley of the exoplanet mass distribution, at $\sim60\unit{M}_\oplus$. From these comparisons it is possible to reaffirm that the outcome of the formation and evolution of CBPs could be different from that of exoplanets around single stars. It is worth mentioning that Kepler-47 is the only multi-planetary circumbinary system, and Kepler-64 (also known as PH1 b) is a circumbinary exoplanet in a four-star system. 
The right panel in Fig.~\ref{fig:m_r_Otegi} shows a zoom onto the region around the CBPs. All CBPs reside above the theoretical $H_2O$ composition line, and therefore could be considered as volatile-rich. These planets may therefore have formed beyond the disc ice line at large semi-major axis before migrating towards the inner disc regions. From this figure, it would appear that binaries with larger eccentricities tend to host larger and more massive planets. A similar correlation seems to be fulfilled when one considers the binary mass instead. By way of comparison, this figure also shows a fit for the mass--radius relationship (valid for volatile-rich planets) presented in OC2020, given by:
\begin{equation}\label{eq:m-r}
    \left(\frac{R}{\mathrm{R}_\oplus}\right) = (0.7 \pm 0.11) \left(\frac{m}{\mathrm{M}_\oplus}\right)^{(0.63 \pm 0.04)},
\end{equation}
where $R$ is the planet radius and $m$ its mass.
Although it is evident that the CBPs do not follow the trend of this fit (red curve in right panel in Fig.~\ref{fig:m_r_Otegi}), it is difficult to calculate a reliable fitting function because of the paucity of  available data.

Figure \ref{fig:R.P_Pb} shows the radii of known CBPs as a function of their period ratio $(P/P_B\equiv n_B/n)$ with their host binaries. None of the observed CBPs have a period ratio of below 5, and most of them have a period ratio of between 5 and 7. It also seems to be the case that the larger the binary eccentricity, the larger the period ratio of the planet. These characteristics seem to be independent of the planet mass. In the following section, we describe our numerical simulations to understand planetary migration around binaries.

\section{Methods}\label{sec:methods}
Based on $n$-body integrations without dissipation forces, we construct dynamical maps in the plane $a-e$ to highlight the location of MMRs and the process of capture when performing the long-term  $n$-body with Stokes-like dissipation forces.
We consider a binary system with masses $m_0$ and $m_1$, respectively, and semi-major axis $a_B=0.1\unit{AU}$. The mass of the planet orbiting the binary is equal to $m=93\unit{M}_\oplus$. We choose a Jacobi system for the coordinates. The orbital elements $a$, $e$, $M$, and $\omega$ are the semi-major axis, eccentricity, mean anomaly, and argument of the pericentre, respectively. We use the subscript $B$ to refer to the binary orbit. The binary and planet orbits are coplanar.

The $n$-body simulations were performed with a Burlisch-Stoer integrator with adaptive step size, which was modified to independently monitor the error in each variable; this imposes a relative precision of better than $10^{-13}$. Integrations are stopped when the distance from the planet to any star is less than the sum of their radii, or when the planet is ejected from the system after scattering between bodies ($>2\unit{AU}$). To construct the dynamical maps (without considering external forces), the total integration time was set to encompass several periods of the orbital secular variations (i.e. $600\unit{yr}\approx20\,000$ binary periods).

The initial conditions for the circumbinary dynamical maps are coplanar, with mean anomaly $M_B=M=0^\circ$ and argument of pericentre $\omega_B=\omega=0^\circ$. To better sample the structure of the resonances, we calculate $\Delta e$ as the amplitude of maximum variation of the orbital eccentricity of the planets during integration, with $\Delta e= \left(e_{\mathrm{max}}-e_{\mathrm{min}}\right)$. This indicator correlates with the chaos indicator {\sc megno} \citep{Cincotta.Simo.1999}, and we verified that higher values of $\Delta e$ (>0.4) correspond to chaotic orbits (i.e. {\sc megno} >2). 

In the dynamical maps, the locations of N/1 resonances were calculated using the method developed by \cite{Gallardo2021}, using a semi-analytical model for planetary MMRs around binary stars. The code used can be found at \href{http://www.fisica.edu.uy/~gallardo/atlas/plares.html}{www.fisica.edu.uy/$\sim$gallardo/atlas/plares.html}. We assume that the planet is aligned with the binary and that the initial mean anomalies of the binary and the planet are equal to zero. We recall that, as shown by \cite{Gallardo2021}, the width of the resonance for coplanar orbits depends on the initial relative value of $\delta\varpi=\varpi_B-\varpi=\omega_B-\omega$. Given that the pericentre of the planet precesses with a short period, the dynamical maps are similar for any initial value of $\delta\varpi$ as we approach the binary. It is also worth mentioning that for resonances closer to the binary, the nominal location of a given MMR depends on the mean anomaly \citep [see e.g. Fig. A.2. in] []{Giuppone2022}

To simulate the disc-induced migration of the planets, we added an ad hoc external force to an N-Body integrator. This additional term behaves as a Stokes non-conservative force:
\be
    \frac{d^2\boldsymbol{r}}{dt^2}=-C(\boldsymbol{v}-\alpha \boldsymbol{v}_c) \,\,,
\ee
where $\boldsymbol{r}$ is the position vector of the planet (in a Jacobian reference plane), $\boldsymbol{v}$ its velocity vector, and $\boldsymbol{v}_c$ the circular velocity vector at the same point \citep{Beauge2006}. With this formulation, the coefficients $C$ and $\alpha$ are defined as
\be
    C=\frac{1}{2\tau_a}+\frac{1}{\tau_e} \,\,, \qquad \alpha=\frac{2\tau_a}{2\tau_a+\tau_e}\,\,,
\ee
where $\tau_a$ and $\tau_e$ are the characteristic semi-major axis and eccentricity damping timescales, respectively.

Following \citet{Beauge1993}, at first order in eccentricity and for a single planet, the effects of the previous force in the semi-major axis and eccentricity of the body can be
described as:
\begin{eqnarray} \label{eq:exponentials}
    a(t) = a_0\exp{\left(-\frac{t}{\tau_a}\right)}\,\,, \qquad
    e(t) = e_0\exp{\left(-\frac{t}{\tau_e}\right)} \,\,,
\end{eqnarray}
where $a_0$ and $e_0$ are the conditions at the beginning of the integration. The $a$-folding and $e$-folding times for $a$ and $e$ are denoted $|\tau_a|$ and $|\tau_e|$, respectively, and have the following expressions:
\begin{eqnarray} \label{eq:variables}
     \tau_a^{-1} = 2C(1-\alpha)\,\,, \qquad
     \tau_e^{-1} = C\alpha\,\,.
 \end{eqnarray}
Following this formalism, the accelerations from tides were incorporated into our $n$-body code, as in \citet{Ronco2020} for other external forces.

Some authors set the initial parameters $\tau_a$, $\tau_e$ from the binary and disc properties \citep{Zoppetti2018, Secunda2019, Secunda2020, Martin2022}, and then evolve them through simulations according to some defined relation \citep{Cresswell2008}. These relations were obtained mainly from hydrodynamical simulations of discs around single stars, and their application to circumbinary planets should be done with caution \citep{Chrenko2018}. 
In order to parameterise $\tau_e$ through simulations, some authors use the simple relation 
\begin{equation}
    \tau_e=\frac{\tau_a}{K}\,\,, 
\end{equation}
where $K$ is a constant with a typical value $K=10$ \citep{Lee2002, Rein2012, Martin2022}.

In our work, we extend the parametric study presented by \citet{Martin2022}, increasing the range of explored parameters. This approach is mainly motivated by the fact that the torque equations around binary systems have not yet been fully resolved; especially near the binary baricentre where important MMRs like 5/1, 4/1, and 3/1 also play an important role in the dynamics. Instead, for our simulations, we set the initial values of $\tau_a$ and $\tau_e$ independently of each other, and they remain constant throughout all simulations.

We vary the following parameters in our simulations: the binary eccentricity and mass ratio ($e_B$, $q$), the mass and initial eccentricity of the planet $(m,\, e)$, and the characteristic timescales of the semi-major axis and eccentricity dampings $(\tau_a,\,\tau_e)$. In all our simulations, the initial semi-major axis of the planet is set at $a=1.5\unit{AU}$, which is sufficiently far from the binary star, which interacts both with the planet and the disc. 

Table \ref{tab:params} summarises the different values adopted for each parameter in our simulations.
Due to the fact that the expected lifetime of each simulation is roughly $\min{(\tau_a,\tau_e)}$, we set the output times for each simulation as follows:
\begin{equation}\label{eq:dtout}
    dt_{\rm out} = 0.5\times10^{\log_{10}(\min(\tau_e,\tau_a)/\unit{yr}) - 2} \unit{yr}\,\,.
\end{equation}
With this configuration, we are able to avoid storing large amounts of output data without subsampling.

\begin{figure*}
    \includegraphics[width=1.\textwidth]{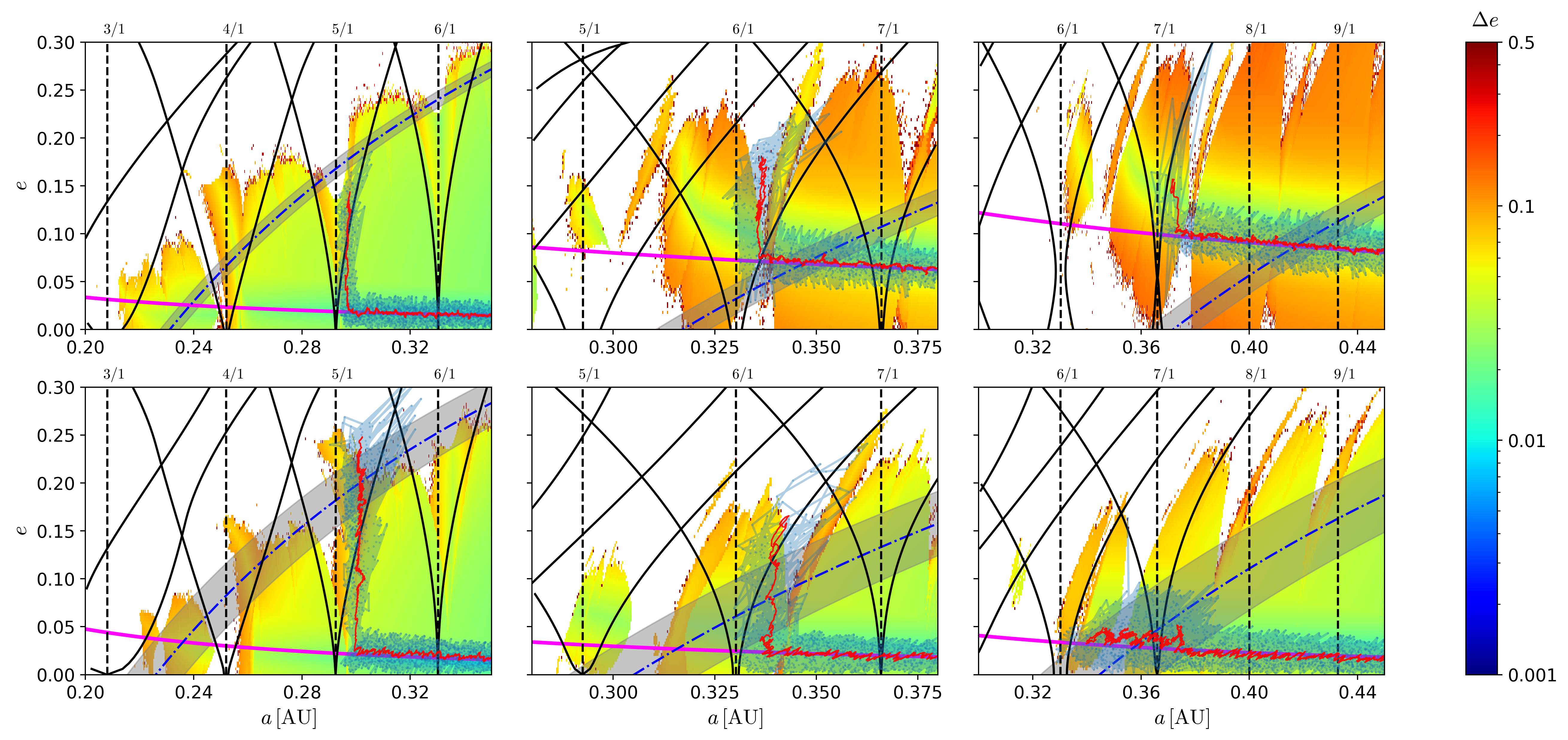}
    \caption{Dynamical maps in the plane $(a,e)$, where each initial condition is integrated for $600\unit{yr}$ and the colour scale corresponds to the $\Delta e$ indicator. Solid black lines denote the width of the most prominent $N/1$ MMR and the dashed vertical lines indicate their nominal location, both calculated using \cite{Gallardo2021}. For each of these maps, the binary has $q=0.2$ (top frames) and $q=1$ (bottom frames), and $e_B=0.05$ (left frames), $e_B=0.3$ (centre frames),  and $e_B=0.5$ (right frames). The grey shaded area denotes the stability limit (the blue dot-dashed line being the reference value) calculated by \cite{Holman+Wiegert1999}, with its critical eccentricity approximated by \cite{Quarles2018}. The solid magenta lines denote the approximation of the `capture' mean eccentricity \citep{Zoppetti2019}. The light-blue dotted line denotes the simulated $(a,e)$ evolution of a planet with $m=5\unit{M}_\oplus$ and $e=0.01$, setting $\tau_a=10^5\unit{yr}$ and $\tau_e=10^6\unit{yr}$. The red solid line denotes the simple moving average of the planet trajectory, with a window of 20 points ($dt_{\rm w}=10^5\unit{yr}$).
    }
    \label{fig:6maps}
\end{figure*}

\begin{table}
\begin{center}
\caption{Parameters adopted in the numerical simulations. }
\label{tab:params}
\begin{tabular}{c|c}
\hline
Parameter & Adopted values\\ \hline
$q$      & 0.2, 1\\
$e_B$    & 0, 0.01, 0.03, 0.05, 0.1, 0.3, 0.5\\
$e$      & 0, 0.001, 0.01, 0.1\\
$m\,$[M$_\oplus$] &  5, 20, 93\\
$\tau_a\,$[yr] & $10^3$, $10^4$, $10^5$, $10^6$, $10^7$\\
$\tau_e\,$[yr] & $10^3$, $10^4$, $10^5$, $10^6$, $10^7$\\
\hline
\end{tabular}
\tablefoot{Simulations with binary mass ratio $q=m_1/m_0=1$ were set up with star masses $m_\star=\lbrace{1,1\rbrace}\unit{M}_\odot$, while the ones with $q=0.2$ were set up with $m_\star=\lbrace{0.833,0.167\rbrace}\unit{M}_\odot$.
}
\end{center}
\end{table}

Instead of shutting down the external ad hoc force (which would take into account the dissipation of the disc), we kept it active throughout the run. By doing so, the disc does not dissipate. We choose this configuration because uncertainty remains over the estimated lifetime of circumbinary discs \citep{Alexander2012, Shadmehri2018, Ronco2021}, although disc lifetime measured observationally is about $3\times10^6\unit{yr}$ \citep{Ribas+2014}.

\section{Dynamical maps and drag of the planets}\label{sec:maps}


Figure \ref{fig:6maps} shows six dynamical maps in the plane $(a,e)$ around a binary with mass ratio $q=\lbrace0.2, 1\rbrace$ (top and bottom frames, respectively) and eccentricities $e_B=\lbrace0.05, 0.3, 0.5\rbrace$ (left, centre, and right frames, respectively). Each dynamical map consists of a grid of $300\times100$ initial conditions integrated for $600\unit{yr}$. The colour scale varies with $\Delta e$, increasing from blue (regular motion) to red (chaotic motion). This figure also shows the location of the most important $N/1$ MMRs, and the stability limit
\begin{equation}\label{eq:aHol}
\begin{aligned}
    a_{\rm c} = a_B &\left(
     1.60^{+0.04}_{-0.04}
     + 5.10^{+0.05}_{-0.05} e_B
     - 2.22^{+0.11}_{-0.11} e_B^2 \right.\\[.1ex]
     &+ 4.12^{+0.09}_{-0.09} \mu
     - 4.27^{+0.17}_{-0.17} \mu e_B
     - 5.09^{+0.11}_{-0.11} \mu^2 \\[.1ex]
     &\left.
     + 4.61^{+0.36}_{-0.36} \mu^2 e_B^2
    \right),
\end{aligned}
\end{equation}
from \cite{Holman+Wiegert1999}, where $\mu=m_1/(m_0 + m_1)$, and using the critical eccentricity approximation,
\begin{equation}
    e_{\rm c} = 0.8 \left(1 - \frac{a_{\rm c}}{a}\right),
\end{equation}
from \cite{Quarles2018}.
These widely used empirical criteria \citep{Yamanaka2019, Zoppetti2020, Gallardo2021, Martin2022} roughly define the smallest stable orbit around a binary system.


In all six maps, it is possible to identify a minimum eccentricity variation region \citep[\textit{secular mode},][]{Moriwaki2004, Paardekooper2012, Zoppetti2019}. The location of the `capture' mean eccentricity of this region can be approximated by
\begin{equation}\label{eq:ecap}
    \langle e\rangle _{\rm cap}=\sqrt{e_{\rm f}^2 + \epsilon^2 \biggl<{\frac{T}{L^*}\biggr>}},
\end{equation}
where
\begin{equation}
   e_{\rm f}=\frac{5 a_B e_B \left(3 e_B^2+4\right) (m_0-m_1)}{8 a \left(3 e_B^2+2\right) (m_0+m_1)}
\end{equation}
is the forced eccentricity, and $\epsilon^2\langle T/L^*\rangle$ is an eccentricity term that can be approximated with the following analytical expression \citep{Paardekooper2012, Zoppetti2019}:
\begin{equation}\label{eq:eps_zop}
    \epsilon^2\langle T/L^*\rangle = \frac{9}{16}\frac{m_0^2 m_1^2}{(m_0 + m_1)^4}\left(\frac{a_B}{a}\right)^4\left(1 + \frac{34}{3}e_B^2\right).
\end{equation}
The values for Eq. \eqref{eq:ecap} are shown in solid magenta lines in Fig. \ref{fig:6maps}.

Even though the \cite{Holman+Wiegert1999} limit considers the growth of the chaotic region (white area of the maps) for binaries with higher eccentricity or mass ratio, it fails to properly enclose the stable region. For example, in the upper right frame of Fig. \ref{fig:6maps}, there is a roughly stable region (of the secular branch) when $a \in [0.36, 0.42]\unit{AU}$ and $e \sim 0.1$, which is considered unstable by this limit. From Fig. 3, it is also important to note the presence of isolated regions that are not fully unstable ($\Delta e \lesssim 0.3$) near $a \in [0.35, 0.4]\unit{AU}$ and $a \in [0.33, 0.35]\unit{AU}$ in the upper and lower right frames (respectively), also incorrectly catalogued as unstable by this limit.

This figure also shows the migration trajectory of a planet of mass $m=5\unit{M}_\oplus$ with initial eccentricity $e=0.01$, $\tau_a=10^5\unit{yr}$, and $\tau_e=10^6\unit{yr}$. The trajectories are plotted back to the last non-ejected state, and in all six simulations the planet got ejected at $\sim1.5\times10^5\unit{yr}$.
In addition to showing every ($a_i,e_i$) point for a planet, its simple moving average (SMA) is also shown as a red line. This estimator is defined as:
\begin{equation}
   SMA(\vec{X}, N)_j  = \frac{\sum_{i=j}^{j+N} X_i}{N},
\end{equation}
where $N$ is the SMA window. $N=20$ for every SMA calculation in this work.

We generally observe that the planet migrates inwards through the {secular mode}, until some MMR captures it. When capture occurs, the planet eccentricity increases with a small variation in the semi-major axis. For the higher binary eccentricity, or lower binary mass ratio, the {secular mode} is located further away from the binary and has higher eccentricity. This relationship causes the capture of exoplanets at higher eccentricities by higher order MMRs in low $q$ or high $e_B$ binaries. It is worth noting that the planets in the upper and lower right frames of Fig. \ref{fig:6maps} were able to reach the stable islands mentioned previously, migrating through the unstable region in between. Eventually, these planets are ejected (see Appendix~\ref{app:ejectimes}).

\begin{figure}
    \includegraphics[width=1.\columnwidth, clip]{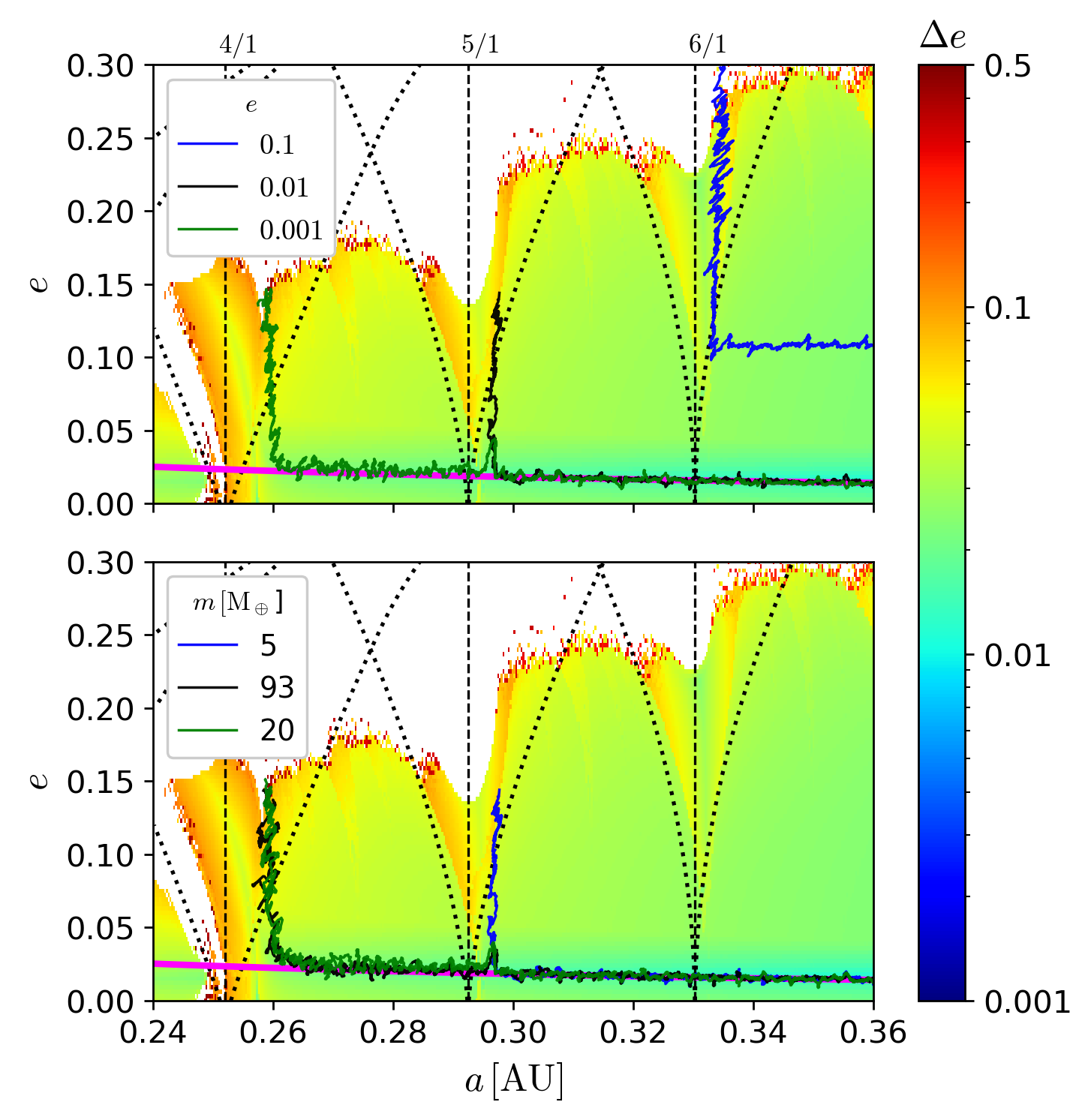}
    \caption{Moving average trajectory in the plane $(a,e)$ of three individual simulations with the same $\tau_a=10^5\unit{yr},\,\tau_e=10^6\unit{yr},\,q=0.2,\,m=5\unit{M}_\oplus, \,$ and $e_B=0.05$, but different initial eccentricity. Vertical dashed lines denote the location of the $4/1$, $5/1,$ and $6/1$ MMRs. The solid magenta lines denote the approximation of the `capture mean eccentricity \citep{Zoppetti2019}.} \label{fig:visual_vary}
\end{figure}

To illustrate the dependence on mass and initial eccentricity of the planet, we show two examples for the same binary ($q=0.2,\, e_B=0.05$) and dissipation parameters ($\tau_a=10^6\unit{yr}$,\, $\tau_e=10^7\unit{yr}$), but setting different initial conditions for the planet. The upper frame of Fig.~\ref{fig:visual_vary} shows the final evolution of three planets with mass $m=5\unit{M}_\oplus$ and initial eccentricities $e=0.001$, $0.01$, and $0.1$. In this case, the planet with eccentricity $e=0.1$ is not able to damp its eccentricity down to the secular branch, and therefore it interacts with the MMR 6/1 within a  region of non-negligible resonance width. As a consequence, the planet is captured by this resonance and subsequently ejected.

On the other hand, planets with eccentricities $e=0.001$ and $e=0.01$ migrate with a low average eccentricity. Even though both of them migrate through the secular branch, the first is captured in MMR 4/1, while the other in MMR 5/1. These results suggest that there is a relationship between the initial eccentricity of the planet and the resonance that captures it. More specifically, it appears that exoplanets with higher initial eccentricity are captured by higher order resonances. We verify that the MMR angle librates when capture occurs. This is shown in Appendix~\ref{app:resangle} for these three specific examples.

The lower frame of Fig.~\ref{fig:visual_vary} shows the final evolution of three planets with eccentricity $e=0.01$ and masses $m=5\unit{M}_\oplus$, $20\unit{M}_\oplus$, and $93\unit{M}_\oplus$. In this case, the mass of the planet also affects the resonance capture. Although it might appear that the results shown in this panel are suggesting that lower mass planets are ejected by higher order resonances, a simple correlation between capture and mass is not evident.

It is not practical to show all our results in this kind of figure because, as mentioned, we explored $4200$ migration simulations. Therefore, in the following section we take a statistical approach to identifying the escape resonance.

\section{Final fate of migrating circumbinary planets}

\subsection{Kernel density estimation for ejecting MMRs}

As shown above, the value of the MMR that captures the migrating planet in each simulation is the most relevant parameter for predicting/understanding its final fate. Given that the ad hoc Stokes force introduced in our simulations is never turned off, it is not surprising to find that the vast majority of exoplanets are eventually ejected by some MMR.
Using a more realistic migration prescription where the Stokes-like force eventually vanishes, we expect planets to remain trapped in those MMRs without being ejected.

Because of the large number of simulations, it is impossible to inspect each simulation carefully to retrieve the MMR that captures and ejects the planet. Therefore, we designed a simple method to calculate this from the trajectory data:
\begin{enumerate}
    \item Calculate the instantaneous semi-major axis ratio $R = \left(a/a_B\right)^{1.5}$ for each output time, which is equivalent to the period ratio $P/P_B$.
    \item Define a Gaussian kernel density estimation (KDE) from the calculated $R$ distribution, with a fixed bandwidth ($=0.03$, in this work).
    \item Calculate the KDE value for fixed bins (points) and retrieve the bin with the highest score.
\end{enumerate}

The $R$ bins (note that this variable is dimensionless) used for these calculations consist of 250 points between $2.5$ and $10.5$ (both included), with a step of $\sim0.032$. This binning contains all possible values of $R$ for our simulations, and allows calculation of the proximity between each $R$ value and its closest MMR. To avoid taking into account the large number of high values of $R$ (corresponding to the first interval of planet migration), we set an upper bound on the semi-major axis for the calculation of the value $R$ at $a_{\max}=1\unit{AU}$.

Multiple results of this method were compared with visual estimates of the trajectories, showing very close agreement. However, because this method is restricted by the $dt_{\rm out}$ of each simulation, it can generate an incorrect classification of $R$ in simulations that do not retain their planet captured in MMR for timescales longer than $dt_{\rm out}$.

\begin{figure}
\centering
\includegraphics[width=1.\columnwidth, clip]{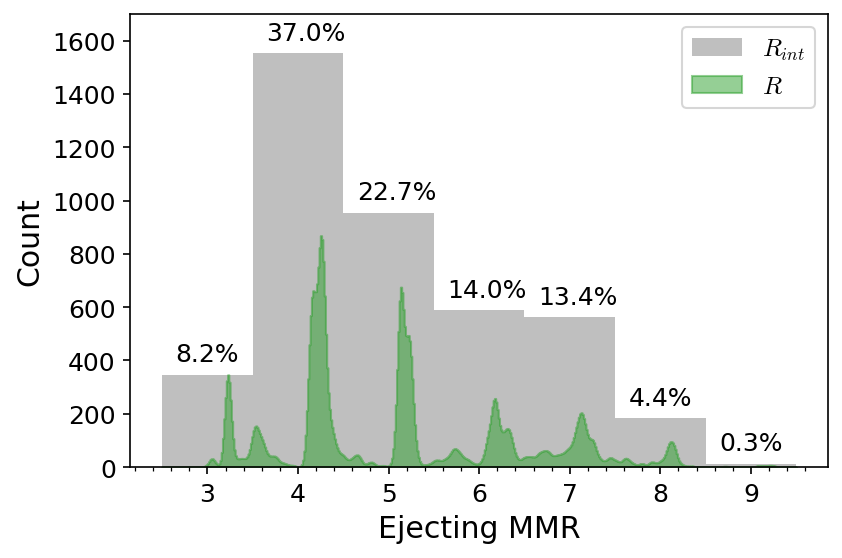}
\caption{Distribution of $R$ values created with KDE, and a histogram of their closest MMR value ($R_{int}$). The values above each $R_{int}$ bin indicate the fraction of the total 4200 simulations found in that bin as a percentage. The $R$ value KDE distribution is re-scaled for improved visualisation.}
\label{fig:Hist_r}
\end{figure}

Figure \ref{fig:Hist_r} shows the distribution of the values $R$ obtained using this method (renormalised for proper visualisation) and the distribution of their closest MMR value. The numbers at the top of each bin indicate the percentage of ejected planets for a given $R_{\rm int}$. Almost all values of $R$ are slightly shifted to the right of their respective MMR nominal location. This result arises from the fact that almost all planets first encounter their ejecting MMR from the right (because they always migrate inward). From this figure, we can confirm that this distribution is far from uniform, and the MMR 4/1 is the one with most planet captures and ejections. 
Of the 4200 planets simulated, only 346 (8.2\%) reached the MMR 3/1, where they were ejected. On the other hand, 1553 of the planets (37\%) were ejected at the MMR 4/1.
For $R \geq 5$, the number of planets ejected decreases with lower MMR values, starting with the minimum value of MMR 9/1, which manages to eject only 12 planets ($<1\%$).

It is worth mentioning that some of the planets with final R values close to the centre between two adjacent MMRs were probably ejected by the contribution of these two MMRs. For example, several $R\sim3.5$ values correspond to planets ejected by the contribution of MMRs 3/1 and 4/1. Generally, we observe that the process of resonant capture is responsible for planetary ejection. 

To establish the final fate of the planets in all simulations in Table \ref{tab:params}, the relevance of each varied parameter must be understood and categorised. In the following section, we analyse the relationship between the parameters and the resulting $R$ value for each simulation.

\subsection{Relevant parameters for predicting capture by MMRs}

\begin{table}
    \centering
    \caption{Minimum, mean, and maximum index of dispersion of $R$ values, calculated for each feature.}
    \begin{tabular}{c|c|c|c|c}
        \multirow{2}{*}{{\bf Parameter}} & \multicolumn{3}{c|}{{\bf Index of dispersion $D$}} & \multirow{2}{*}{\bf {\makecell[c]{Number of \\ $R$ values}}}\\
                        & {\it Min} & {\it Mean} & {\it Max} & \\ \hline
        $e_B$           & 0.044     & 0.059      & 0.098     & 600  \\ \hline
        $q$             & 0.288     & 0.328      & 0.369     & 2100 \\ \hline
        $m$             & 0.325     & 0.329      & 0.333     & 1400 \\ \hline
        $e$             & 0.307     & 0.321      & 0.336     & 1050 \\ \hline
        $\tau_a$        & 0.275     & 0.325      & 0.368     & 840  \\ \hline
        $\tau_e$        & 0.268     & 0.322      & 0.359     & 840  \\ \hline
        $\tau_a/\tau_e$ & 0.273     & 0.317      & 0.340     & 168  \\ \hline
    \end{tabular}
    \tablefoot{The last column shows the number of simulations used to calculate each index of dispersion $D$.}
    \label{tab:variances}
\end{table}

One method to evaluate the importance of each feature\footnote{In this section we prefer to use the term `feature' instead of `parameter', as this is typical machine learning terminology.} within the simulation dataset is to analyse the index of dispersion ($D$) of a target over all single values of each feature \citep{Cox1966}. This index measures how closely a set of observed occurrences resembles a typical statistical model in terms of clustering or dispersion. This index is defined as the ratio of the variance $\sigma^2$ over the mean value $\mu$: $D=\sigma^2/\mu$. In our case, we take the instantaneous semi-major axis ratio ($R$) as the target value.

Table~\ref{tab:variances} shows the minimum, mean, and maximum index of dispersion of the $R$ values, calculated among the indices obtained for each single value of each parameter in Table~\ref{tab:params}. $\tau_a/\tau_e$ is also considered as an additional (derived) parameter. The distribution of $R$ values for each $e_B$ value has the lowest dispersion index of all the features. Therefore, $e_B$ is the parameter that is the most closely associated with the final value $R$ of each planet. This information is connected to the fact that the binary eccentricity is directly related to the position of the innermost stable MMR, as mentioned in Sect.~\ref{sec:maps} (Eqs. \ref{eq:aHol} - \ref{eq:eps_zop}). The remaining features have indices that are comparable to each other, and so it is not possible to clearly determine the order of their respective importance with this method.

\begin{figure}
    \centering
    \includegraphics[width=1.\columnwidth, clip]{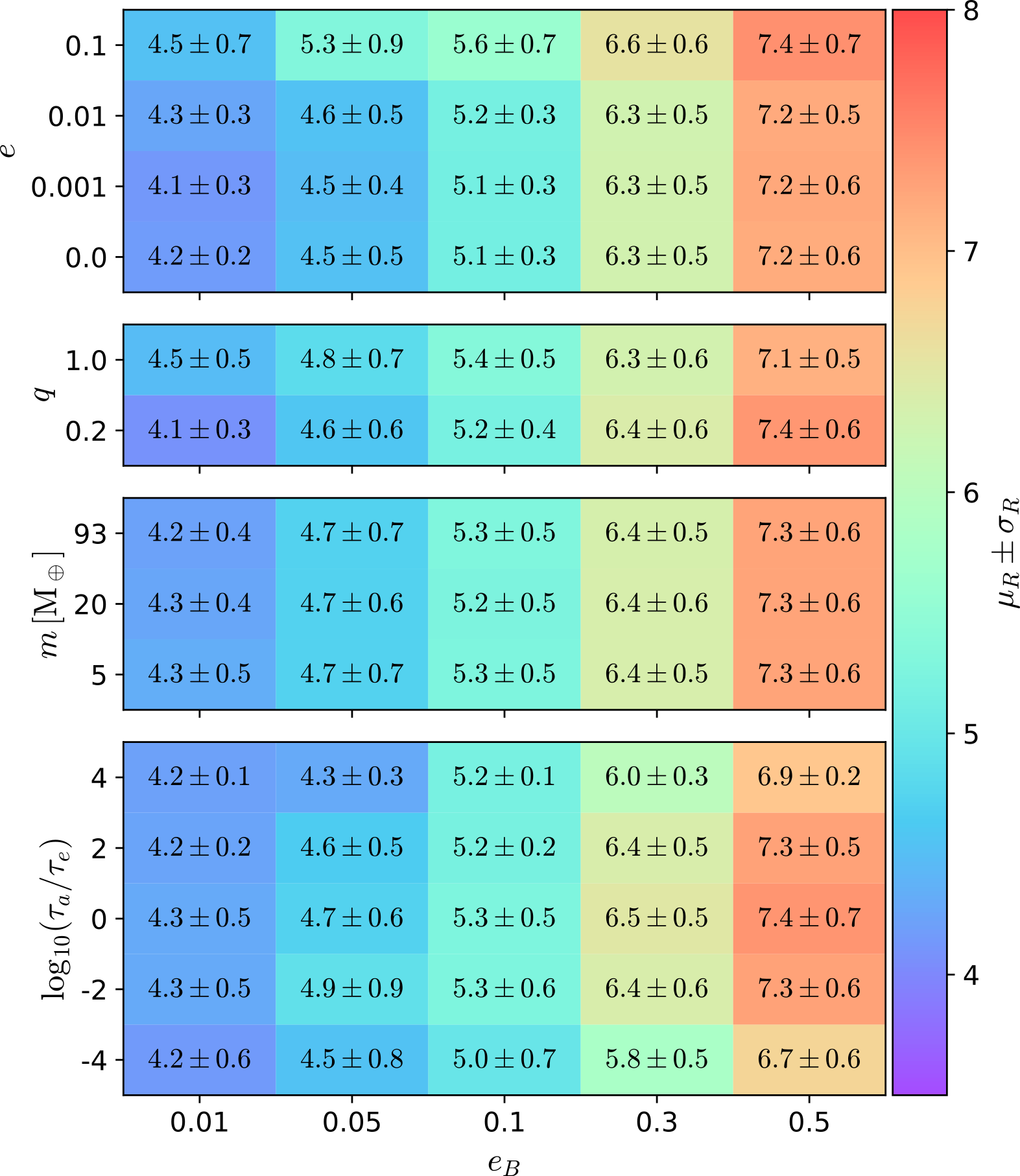}
    \caption{Average values of $R$ obtained for simulations characterised by $e_B$.
    The y-axis denotes the projection of all the simulations depending on $e$, $q$, $m,$ and $\log_{10}(\tau_a/\tau_e)$, from top to bottom panels, respectively.}
    \label{fig:all_grids}
\end{figure}

To visualise the effect of varying each parameter of Table~\ref{tab:params}, we present {average resonance tables} in Fig.~\ref{fig:all_grids}. This figure shows the effect of different initial $e$ (top frame), $q$ (second frame), $m$ (third frame), and $\log_{10}(\tau_a/\tau_e)$ (bottom frame) on the average $R$ value ($\mu_R=\bar{R}$) and its dispersion ($\sigma_R$). All grids have $e_B$ as their primary axis because of its strong relationship with $R$.

From the top panel, we can determine that planets with higher $e\,(=0.1)$ tend to be captured by slightly higher order MMRs (ranging from 4.5 to 7.4). This fact is related to the resonant capture of planets with high eccentricity, which is discussed at the end of Sect.~\ref{sec:maps}. There are no appreciable differences in the values of $\mu_R$ between planets with low eccentricity ($e<0.1$).

From the second panel, it is possible to infer that the dependence of $\mu_R$ on $q$ is weakly related to the binary eccentricity. For a large value of the mass ratio, the increase rate of $\mu_R$ with respect to $e_B$ has a smaller slope compared to systems with a smaller mass ratio. In this case, the slope for simulations with $q=1$ is $\sim(5.27\pm0.48)\,\mu_R/e_B$, while that for simulations with $q=0.2$ is $\sim(6.49\pm0.60)\,\mu_R/e_B$.

From the third panel, the value of $\mu_R$ does not vary as the mass of the planet changes. Therefore, regarding the capture in MMRs, the planetary mass is a parameter of minor significance.
From the lower panel it is possible to determine the relationship between $\log_{10}(\tau_a/\tau_e)$ and $\mu_R$. The value of $\mu_R$ decreases as the value of $|\log_{10}(\tau_a/\tau_e)|$ increases, regardless of whether $\tau_a$ is higher or lower than $\tau_e$. This effect is more noticeable as the binary eccentricity increases. It is interesting to note that, although the values of $\mu_R$ are similar, their dispersion values $\sigma_R$ are slightly higher for the simulation groups with $\tau_a\le\tau_e$.

\subsection{Methods to predict the MMR at which ejection occurs}

The large number of simulations performed allowed us to carry out a statistical analysis to determine whether it is feasible to predict the MMR at which a CBP will be caught, depending on the initial values of the parameters examined. For this purpose, we implemented a simple random forest (RF) model.

As an ensemble learner, a RF generates a large number of classifiers and aggregates their results.
This algorithm builds several classification (or regression) trees, each of which will be trained on a bootstrap sample of the initial training data and will search on a randomly chosen subset of input variables to evaluate the splitting \citep{Breiman2001}. For a classification task, the class chosen by the largest number of trees is the RF output. To predict a variable, this predictive modelling tool pulls information from input data sets and seeks to identify new associations \citep{Ivezic2014}.
In this work, the closest nominal MMR of an exoplanet is predicted by the classification task of this tool given initial condition parameters as input.
For this task, we used the algorithm {\tt RandomForestClassifier}\footnote{\href{https://scikit-learn.org/stable/modules/generated/sklearn.ensemble.RandomForestClassifier.html}{https://scikit-learn.org/stable/modules/generated/ \newline sklearn.ensemble.RandomForestClassifier.html}} from the {\tt scikit-learn} Python package.

From the sample of 4200 simulations, we designed a training set made up of 75\% of them (3150) and a test set made up of the remaining 25\% (1050). The data columns (input parameters) of each set are $\left[ e_B,\, e,\, \tau_a,\, \tau_e,\, q,\, m,\, \tau_a/\tau_e \right]$, and their target is the integer closest to each value $R$, named $R_{\rm int}\equiv MMR$. For this simple random forest classifier (RFC) model, we used the default hyper parameters from {\tt scikit-learn}.

\begin{figure}
\centering
\includegraphics[width=1.\columnwidth, clip]{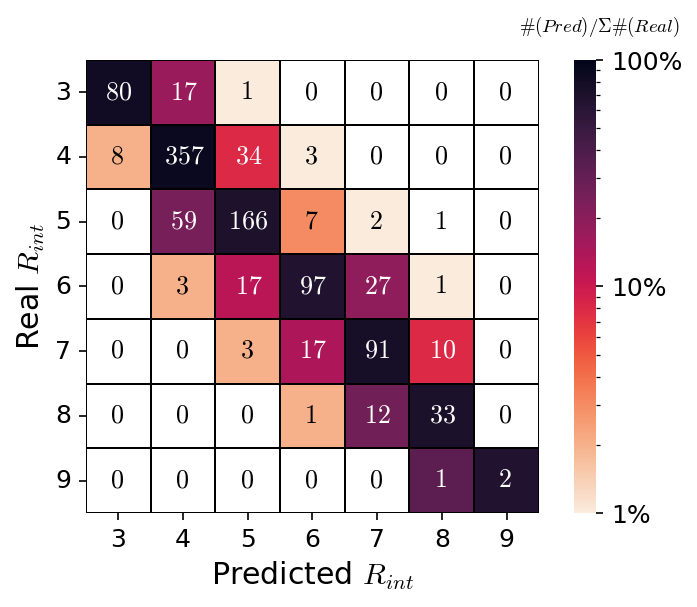}
\caption{Confusion matrix of the RFC used, applied to the test data set. The colour bar denotes the percentage that each predicted amount represents normalised to the amount of real values of each label.}
\label{fig:conf_mat}
\end{figure}

Figure \ref{fig:conf_mat} and Table \ref{tab:RFC-metrics} show the confusion matrix and a classification report (respectively) of our classifier applied to the test data set. This data set was never seen by the respective estimator during training. The metrics presented here are precision ($TP/(TP+FP)$), recall ($TP/(TP+FN)$), and F1 score ($2\times {\rm precision}\times {\rm recall} / ({\rm precision} + {\rm recall})$), where $T(F)P(N)$ are the number of true (false) positive (negative) values. These metrics are some of the most widely used to measure the general performance of a trained model \citep{Bengfort2019, Schlecker2021, Audenaert2021}. Table \ref{tab:RFC-metrics} also shows the support for each class, which is the actual number of instances of each class in the test data set.

The precision rate, which can be seen as a measure of the exactness of a classifier, is greater than 70\% for all classes (except for $R_{\rm int}=7$, which has a rate equal to 69\%). Moreover, the accuracy score of the entire set, calculated as the weighted average (averaging the support-weighted mean per label) of the precision rates, is 79\%. Although this metric indicates how often the model makes a correct prediction, given the severely unbalanced nature of the data set, it is insufficient as a performance indicator. Therefore, we assess our model's performance and examine the resulting different kinds of errors.

We can see in Figure \ref{fig:conf_mat} that there is a significant difference between the colors of the diagonal and those of the off-diagonal entries, which indicates that our model has good performance. To retrieve the completeness of the classifier, which can be described as its ability to correctly classify all true instances of a given class, the recall rate is measured. Table \ref{tab:RFC-metrics} shows that, except for $R_{\rm int}=6$ and $R_{\rm int}=9$, all classes have a recall rate of greater than 70\%, with 89\% being the highest. The highest misclassification rate (33\%) occurred between classes 8 and 9 (as one of the only three instances of $R_{\rm int}=9$ was classified as $R_{\rm int}=8$), and the highest accuracy rate (91\%) was obtained for class 3. It is interesting to note that of all misclassifications, the vast majority occurred between adjacent classes (e.g. $R_{\rm int}=4$ with $R_{\rm int}=5$, or $R_{\rm int}=8$ with $R_{\rm int}=7$). This result indicates that the classifier was able to generate close relationships between the parameter set and the target classes.

\begin{table}
    \centering
    \caption{Classification report of the RFC used.}
    \begin{tabular}{c|c|c|c|c}
    $\boldsymbol{R_{int}}$ & {\bf Precision} & {\bf Recall} & {\bf F1-score} & {\bf Support} \\ \hline
    3 & 0.91 & 0.82 & 0.86 & 98 \\ \hline
    4 & 0.82 & 0.89 & 0.85 & 402 \\ \hline
    5 & 0.75 & 0.71 & 0.73 & 235 \\ \hline
    6 & 0.78 & 0.67 & 0.72 & 145 \\ \hline
    7 & 0.69 & 0.75 & 0.72 & 121 \\ \hline
    8 & 0.72 & 0.72 & 0.72 & 46 \\ \hline
    9 & 1.00 & 0.67 & 0.80 & 3 \\ \hline\hline
    {\bf \makecell[c]{Macro \\ average}} & 0.81 & 0.75 & 0.77 & 1050 \\ \hline
    {\bf \makecell[c]{Weighted \\ average}} & 0.79 & 0.79 & 0.79 & 1050 \\ \hline
    \end{tabular}
    \label{tab:RFC-metrics}
\end{table}

\begin{figure}
\centering
\includegraphics[width=1.\columnwidth, clip]{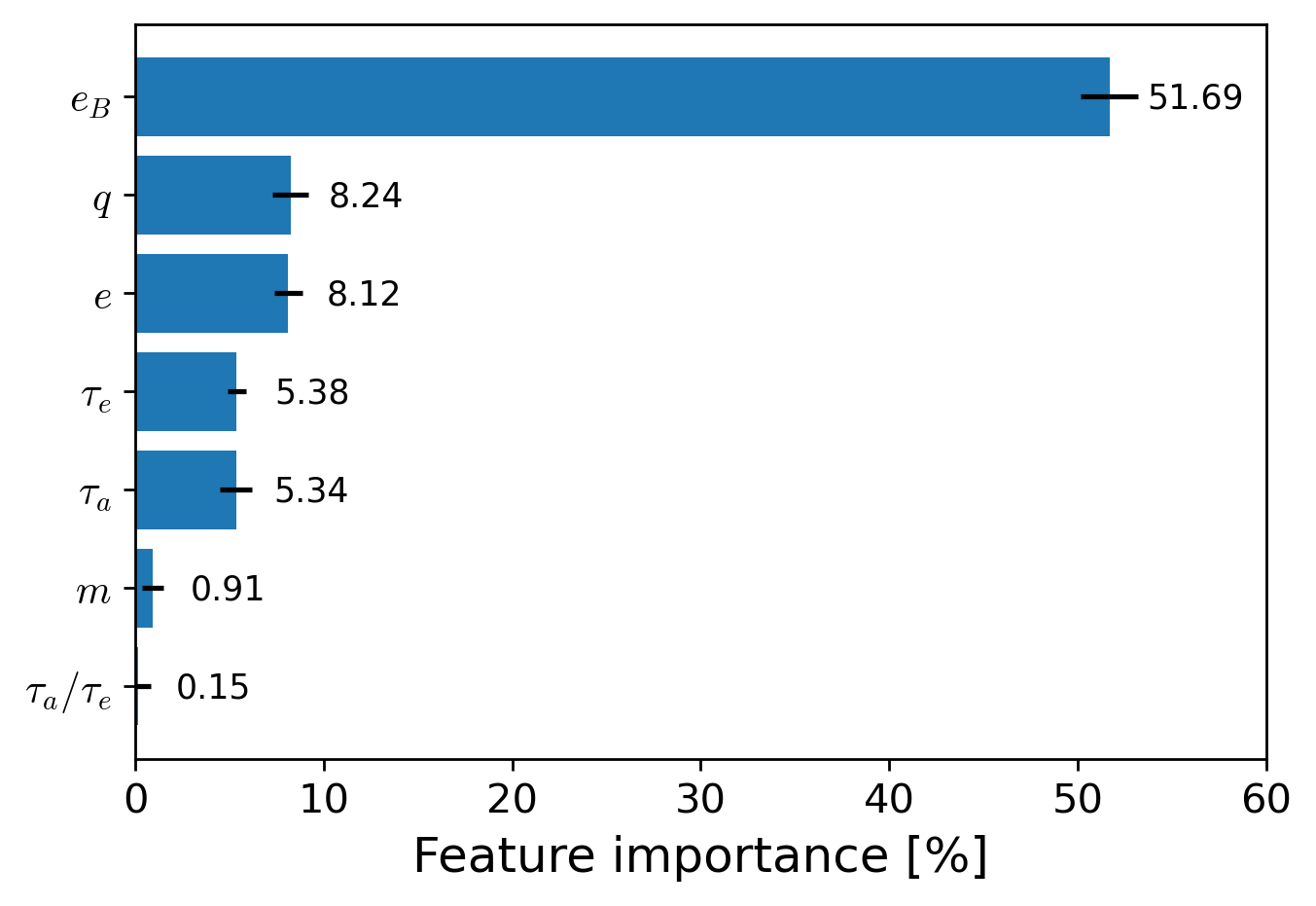}
\caption{Importance percentage of each feature according to our RFC. The black horizontal lines denote the standard deviation of each estimate.}
\label{fig:feat_imp}
\end{figure}

Using this model, it is not possible to establish a parametric relationship between $R_{\rm int}$ and the input parameters \citep{Ulmer-Moll2019}. However, we are able to measure the importance of each feature of the data set according to the classifier. `Permutation feature importance' is a technique for inspecting any fitted estimator when the data are tabular. This method retrieves the drop in model performance caused by randomly shuffling a single feature value \citep{Breiman2001}. This procedure demonstrates the extent to which the model depends on the feature by breaking the feature--target relationship \citep{Lu2020}. The benefit of this method is that it can be calculated numerous times with various feature permutations and is model-independent. Figure~\ref{fig:feat_imp} shows the importance of trained RFC characteristics, which can be understood as the contribution of each individual parameter to the final prediction. These values were calculated using permutation feature importance, with $15$ permutations.

With almost $52\%$  importance, the eccentricity of the binary is the most sensitive feature according to the classifier. All of the other parameters have less than $10\%$ importance, with the stellar mass ratio and initial eccentricity of the  planet being the second-most and third-most important ($\sim8.2\%$), respectively. $\tau_a$ and $\tau_e$ are the next two parameters in terms of importance, both with $\sim 5.3\%$. The planet mass and the ratio of the characteristic damping times are the two least important features with less than $1\%$  importance. The low importance of the time ratio feature is likely due to the fact that the properties that this feature represents in the $R_{\rm int}$ distribution end up being redundant if the properties represented by $\tau_a$ and $\tau_e$ have been previously analysed by the classifier.

\begin{figure}
\centering
\includegraphics[width=1.\columnwidth, clip]{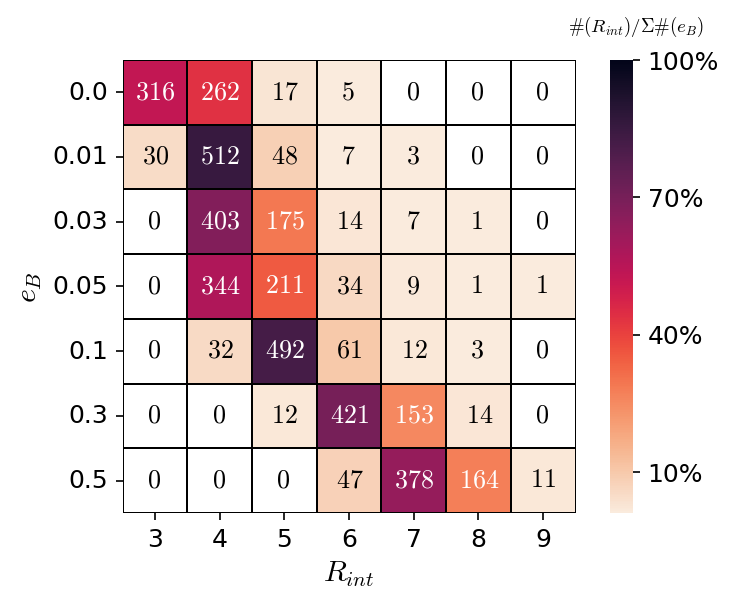}
\caption{Grid with the number of exoplanets according to the binary eccentricity and the final $R_{\rm int}$. The colour bar denotes the normalised percentage that each quantity represents out of the total number of simulations for each eccentricity value (600).} \label{fig:be_rint}
\end{figure}

Considering the high importance of the binary eccentricity with respect to the output MMR of each exoplanet, it is necessary to analyse the distribution of $R_{\rm int}$ with respect to this parameter. Figure \ref{fig:be_rint} shows the number of simulated exoplanets ejected by each $R_{\rm int}$ parameterised by $e_B$. The colour scale shows the percentage representation of each value with respect to the 600 simulations performed for each $e_B$. From this figure, we note that the higher the binary eccentricity, the higher the value of the predominant capture MMR. Furthermore, it is interesting to note that the label $R_{\rm int} = 4$ is mainly obtained for simulations with $e_B=\lbrace0.01,\, 0.03,\, 0.05\rbrace$, while each remaining $R_{\rm int}$ has only one principal binary eccentricity associated with it.


\section{Discussion}\label{sec:discussions}

According to the current paradigm, planets do not form {in situ} around binary stars, but rather migrate from farther out in the disc, where pebble accretion is more favourable. Planets are expected to migrate at a speed proportional to their mass. Remarkably, a slower migration rate makes resonant capture and subsequent ejection more likely.

\citet{Thommes2003} found that two mutually inclined planets migrating in a disc can be captured in resonant inclination when the planetary mass of the  outer planet is greater than half that of the inner planet.
In our model three-body system, the mass of the  planet is negligible in comparison to that of the binary companion. Therefore, we did not consider this resonance configuration in the present work. In principle, it is possible to include an extra term to account for the inclination damping as follows: $i(t) = i_0\exp{\left(-t/\tau_i\right)}$, but this requires additional parameterisation. This specific aspect is left for future work.

\citet{Martin2022} showed that while large planets (roughly $>20\unit{M}_\oplus$) may be able to cross the resonances successfully, small planets (roughly $<3\unit{M}_\oplus$) are prone to being captured. These authors considered analytical prescriptions for $\tau_a$ and $\tau_e$ that modelled Lindblad torques, co-orbital torques, disc surface density, circular gaps in the disc, and stochastic forces to account for the disc turbulence, and demonstrated that the process of resonant ejection of migrating planets may occur in nature. In particular, this mechanism preferentially affects small planets, but is not sufficient to fully explain the dearth of $R_p < 3 \unit{R}_\oplus$ planets.

Nevertheless, there is likely an important observational bias against the co-existence of small planets due to photometric limits \citep [see discussion by][]{Martin2014}.

In this work, we extend the scope of the study by \citet{Martin2022}, because their analytical results were developed for larger distances from the binary barycentre ($a/a_B\gg 1$). We explored a wider variety of binary eccentricities and damping times. In this sense, our models are compatible with a broader range of circumbinary discs. We find that the 4/1 MMR is the most efficient for resonant capture. However, other resonances can also play a significant role in determining the final parking location of circumbinary planets.

\begin{figure}
    \centering
    \includegraphics[width=1.\columnwidth, clip]{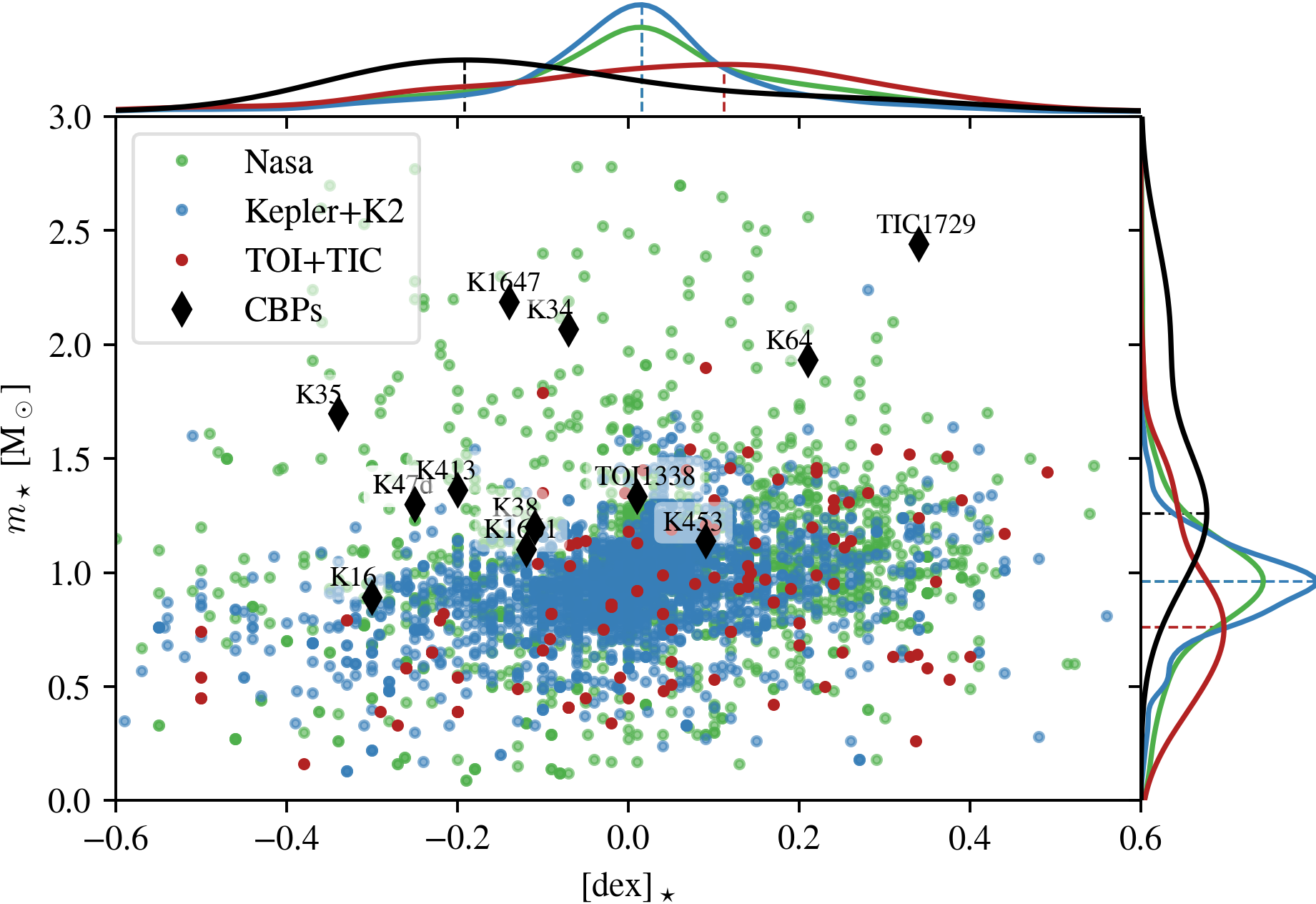}
    \caption{Metallicity of stars harbouring known exoplanets compared with the mass of the host star. For binary stars, we assume the total mass ($m_\star=m_B$). We identify the groups corresponding to all planets, planets detected by Kepler (inc. K2), and TOI (including TIC) planets. Exoplanets data obtained from Nasa Exoplanet Archive database.}
    \label{fig:meta}
\end{figure}

More importantly, circumbinary discs generally form eccentric gaps with a constant precession rate \citep{Thun2018, Kley2019, Penzlin2020} that may alter the torques from the disc. 
Surface density ($\Sigma$), viscosity ($\nu$),  disc turbulence ($\alpha$), and disc metallicity may determine the exoplanet incidence \citep[see e.g.][]{Adibekyan2019}. As shown in Figure~\ref{fig:meta}, the stellar metallicities of detected CBP systems are statistically lower than the metallicities of single stars harbouring planets. This strongly suggests that we may need to reconsider the {typical} disc properties around binary stars. To compare with the sample of CBPs, we show only exoplanets around single stars with a radius of less than $12\unit{R}_\oplus$ and periods of less than $300\unit{days}$. For more realistic discs than the ones considered here, it is likely that the process of MMR capture occurs in any case. Nevertheless, one should not disregard the possibility that, as the planet gains mass and approaches the binary, it switches to type II migration \citep{Armitage2020}. Also, the gas is expected to dissipate after a few million years, which would translate into vanishing disc torques. Lastly, other mechanisms (such as turbulence or mass accretion) might help to avoid the ejection of circumbinary planets.

\section{Conclusions}\label{sec:conclusions}
In this work, we analyse the migration of planets with $5$, $20$, and $93\unit{M}_\oplus$ using Stokes-type forces that mimic planetary migration with a constant rate. Using the mass--radius relation presented in Eq. \eqref{eq:m-r}, we estimate that their radii are approximately $1.9$, $4.6$, and $12.2\unit{R}_\oplus$ (respectively). These values are close to the radius distribution of known CBPs (see Table~\ref{tab:known_planets}).
 
As circumbinary disc dissipation timescales are not yet fully understood \citep{Alexander2012, Shadmehri2018, Ronco2021}, we considered wide ranges of $a$-folding and $e$-folding times, which remain constant throughout the simulation. We also considered a wide range of binary eccentricities. Analysing the final fate of the simulated planets, we find that:
\begin{itemize}
    \item Circumbinary planets generally migrate along a secular branch until they reach some MMR. At this secular branch, the angle $\Delta\varpi$ usually librates.
    \item The vast majority of CBPs are ejected at MMRs when migrating towards the binary, with the MMR 4/1 being the one generating the greatest number of ejections.
    \item The resonant captures always show the resonant angle librating, mainly around 180 degrees ($\psi_{p/q}^{(l)}=(p+q)\lambda - p\lambda_B - q\varpi - l(\varpi_B - \varpi)$ where $q=1$ and $l=1$).
\end{itemize}

Using ensemble learning methods for classification, we draw the following conclusions:
\begin{itemize}
    \item The binary eccentricity is the most important parameter for determining the (lowest order) MMR at which a CBP may be captured. On the other hand, the planetary mass has low significance.
    \item It is possible to predict the lowest $R_{int}$ (defined as the closest integer to the feature $R$, and associated to the nearest MMR) at which a CBP would be ejected, with a precision of $\sim 81\%$. To do so, six parameters are required: the binary eccentricity and mass ratio, the planet eccentricity and mass, and the semi-major axis and eccentricity damping times.
\end{itemize}

Over the last few years, this topic has become an active field of research in planet formation. Indeed, while finishing this study, \cite{FitzmauriceMartinFabrycky2022} studied the migration of two circumbinary planets using the approximation of torques induced by a single star. Improvements in our model can be made by assuming some recipes for planetary accretion, which could be important in the runaway regime for large planets ($>20\unit{M}_\oplus$) and disc dissipation timescale. The development of more refined models is left for a forthcoming work. In any case, it is of crucial importance to accurately establish how planets migrate within circumbinary discs. This will help us to understand where and for how long circumbinary planets are able to survive around stellar binaries.

\begin{acknowledgements}
N-Body computations were performed at Clemente Cluster from IATE, Argentina and at the Mulatona Cluster from the CCAD-UNC, which is part of SNCAD-MinCyT, Argentina. This project has received funding from the European Union’s Horizon 2020 research and innovation programme under the Marie  Sk\l{}odowska-Curie grant agreement No 896319 (SANDS). This research was funded, in part, by ANR (Agence Nationale de la Recherche) of France under contract number ANR-22-ERCS-0002-01. This project has received funding from the European Research Council (ERC) under the European Union Horizon 2020 research and innovation program (grant agreement No. 101042275, project Stellar-MADE).
\end{acknowledgements}
    
\bibliographystyle{aa}
\bibliography{migrationbinaries}

\begin{appendix}

\section{Average ejection times}\label{app:ejectimes}
\begin{figure}[h]
\centering
\includegraphics[width=0.98\columnwidth, clip]{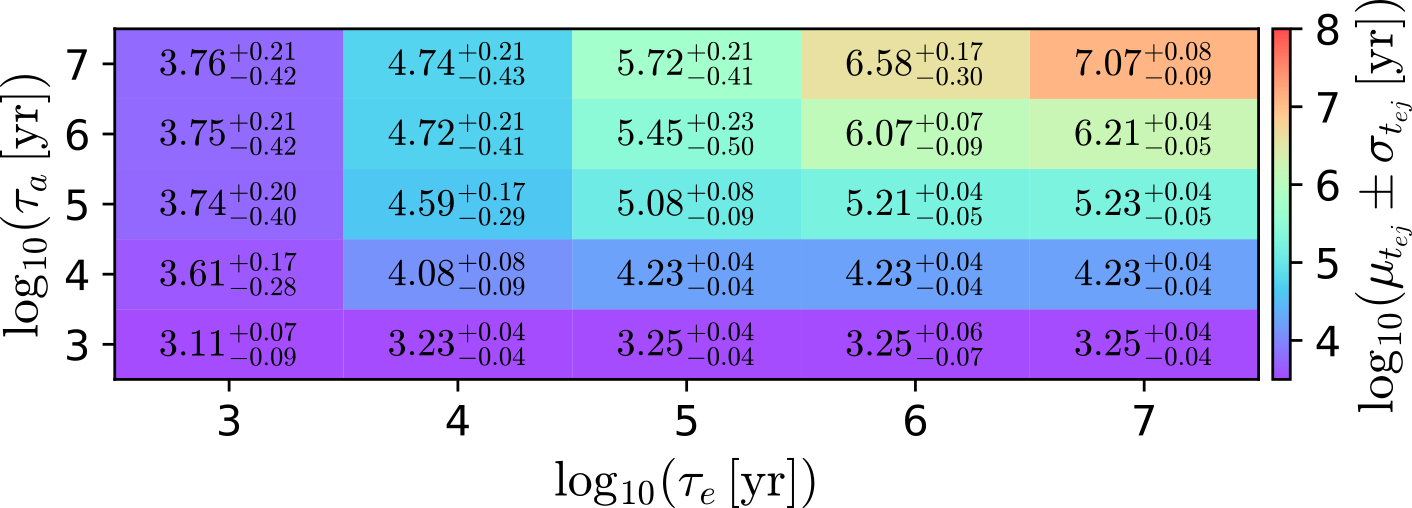}
\caption{Logarithm of the average ejection times ($\log_{10}(\mu_{t_{ej}})$) obtained for simulations, characterised by $\tau_a$ and $\tau_e$. The cells $(\tau_a=10^6\unit{yr},\tau_e=10^5\unit{yr})$ do not take into account the three non-ejected planets.}\label{fig:tauxtimes}
\end{figure}

\begin{figure}
\centering
\includegraphics[width=0.98\columnwidth]{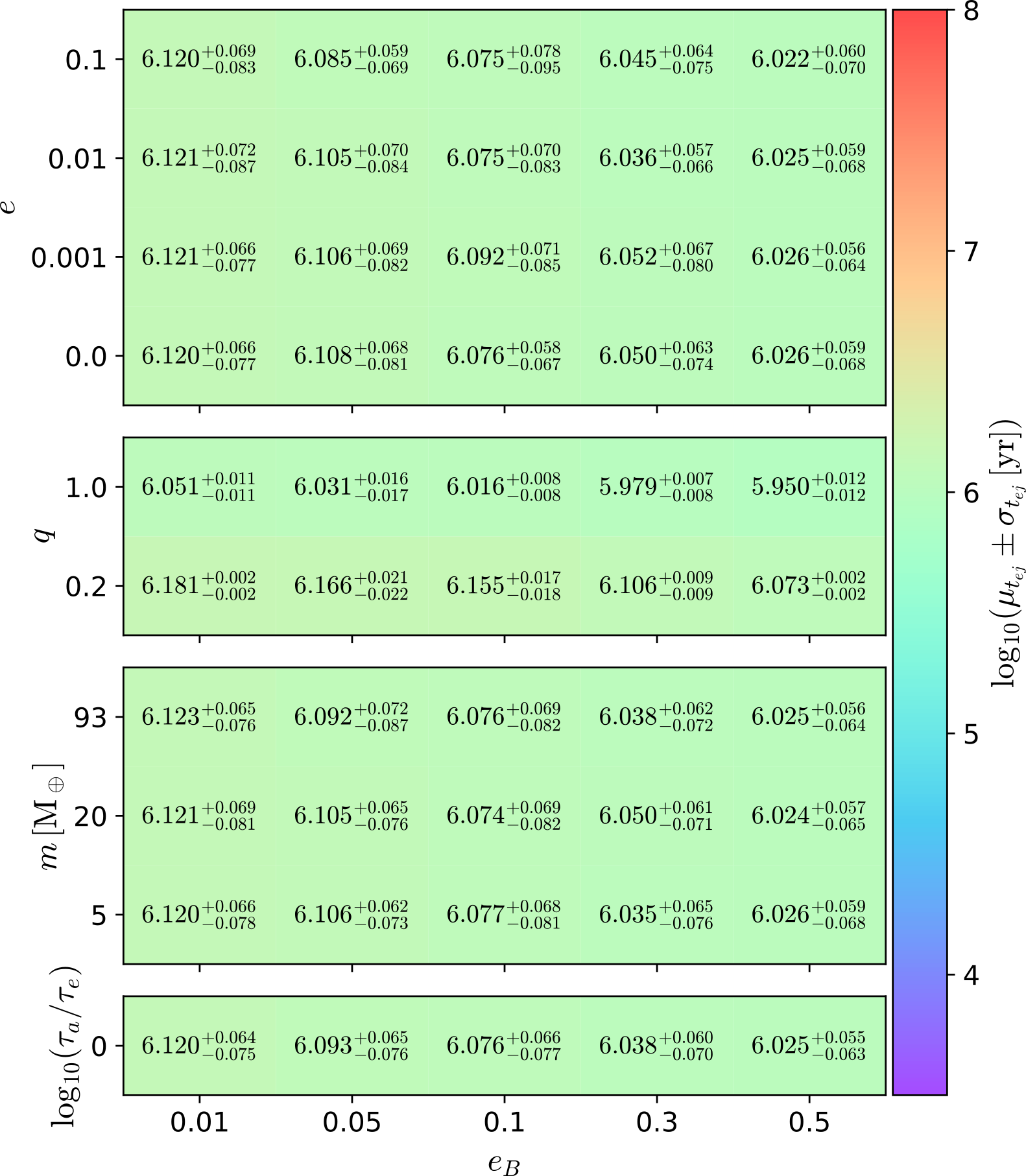}
\caption{Logarithm of average ejection times obtained from simulations, characterised by $e_B$. The y-axis denotes the projection of all the
simulations depending on $e$, $q,$ and $\log_{10}(\tau_a/\tau_e)$, from top
to bottom panels, respectively. In this figure, only the case $\tau_a=\tau_e=10^6\unit{yr}$ is shown.}\label{fig:single_times}
\end{figure}

Figure \ref{fig:tauxtimes} shows a colour scale representing the logarithm of the average ejection times of all simulations in the plane $(\tau_a,\tau_e)$. For a specific  pair of folding times, the values represent the average ejection time and its dispersion within our simulations. We can see that $t_{ej}$ increases as $\tau_a$ and/or $\tau_e$ increase(s). In particular, as mentioned in Sect. \ref{sec:methods}, the mean survival time for each simulation is $\sim\min(\tau_a,\tau_e)$. It is worth mentioning that the errors for this parameter are quite small, giving confidence to the estimates. For example, in $(\tau_a,\tau_e)=(10^6,10^6)\unit{yr}$, all simulations, in general, are ejected in $\sim10^{6.07^{+0.07}_{-0.09}}\unit{yr}$. To corroborate this result, we present Fig. \ref{fig:single_times}. This figure is analogous to Fig. \ref{fig:all_grids}, but shows the averages of $t_{ej}$ instead of $R$, and only for the simulations with $\tau_a=10^6\unit{yr}$ and $\tau_e=10^6\unit{yr}$. All the values are $\sim t_{ej}=10^6\unit{yr}$. 

\section{Resonant angle}\label{app:resangle}
\begin{figure*}
\includegraphics[width=\textwidth]{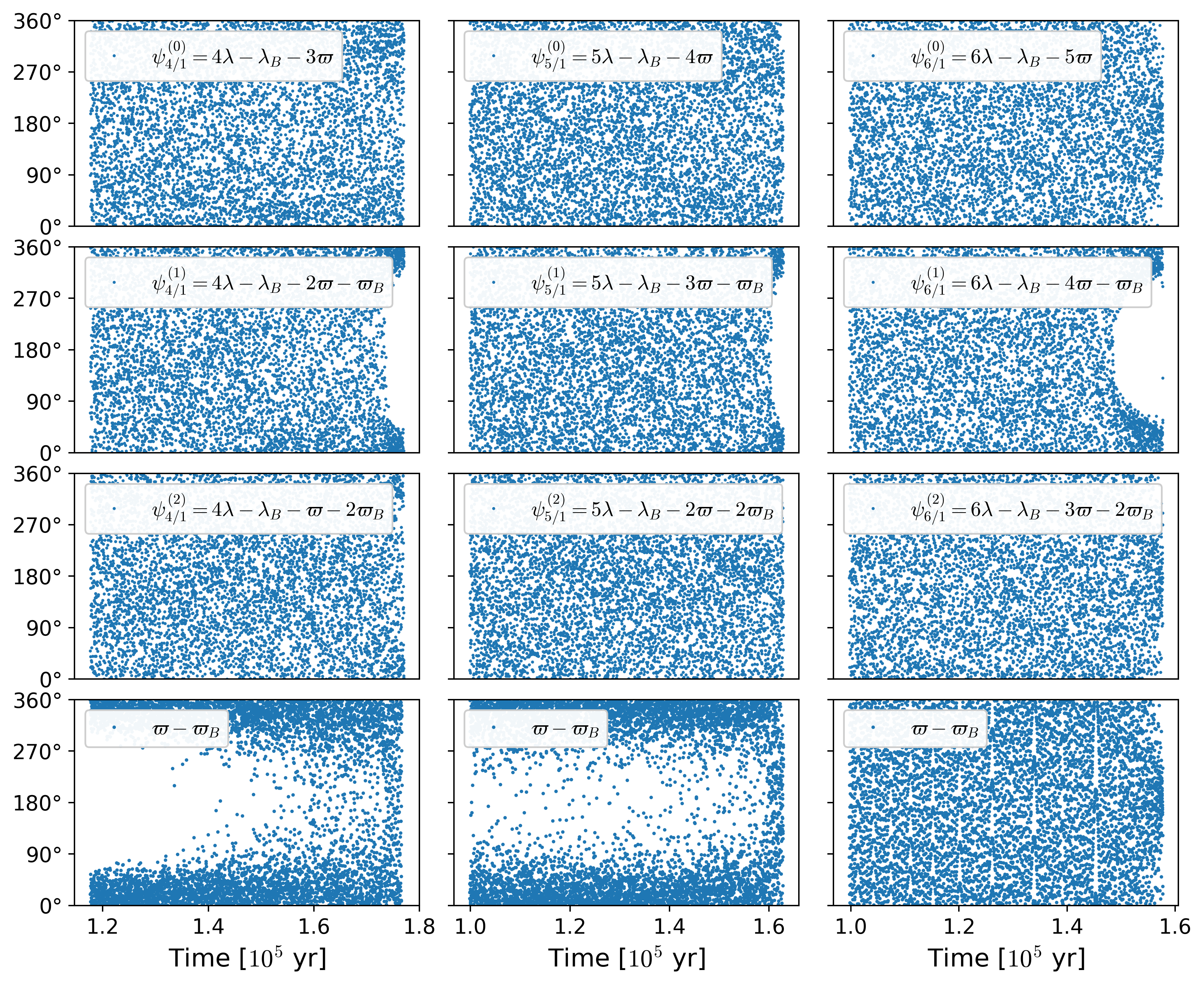}
\caption{Resonant angle variation $\psi_{p/q}^{(l)}=(p+q)\lambda - p\lambda_B - q\varpi - l(\varpi_B - \varpi)$ of the three simulations presented in the top panel of Figure \ref{fig:visual_vary}. The eccentricities of the planets are $e=0.001$ (left), $e=0.01$ (centre), and $e=0.1$ (right), respectively. Only the three first resonant angles (of each simulation) are shown. The angle $\Delta\varpi$ is also shown in the last row.}\label{fig:resonangle}
\end{figure*}

Figure \ref{fig:resonangle} shows the variation of the resonant angle ($\psi_{p/q}^{(l)}$) and the secular angle ($\Delta\varpi$) of the numerical simulations presented in Fig.~\ref{fig:6maps}. Here, in the three simulations, we can see  the resonant angle circulating until reaching (and being captured by) the corresponding MMR. From this point, the angle starts to librate. By analysing the angle $\Delta\varpi$, we can determine that exoplanets with initial eccentricities $e=0.01$ and $e=0.05$ migrated while captured by the secular mode ($\Delta\varpi \sim 0$). On the other hand, for the extreme case of the exoplanet with initial eccentricity $e=0.1$, the semi-major axis and eccentricity dampings are not sufficient for it to be captured by the secular branch. Consequently, in this case, $\Delta\varpi$ is continuously circulating.

\end{appendix}

\end{document}